\documentclass[twocolumn,showpacs,showkeys,preprintnumbers,amsmath,amssymb]{revtex4}
\usepackage{graphicx}% Include figure files
\usepackage{dcolumn}% Align table columns on decimal point
\usepackage{bm}% bold math

\begin{document}

\preprint{APS/123-QED}

\title{Unifying Evolutionary and Network Dynamics}

\author{Samarth Swarup}
 \email{swarup@uiuc.edu}
\affiliation{
Department of Computer Science,\\
University of Illinois at Urbana-Champaign
}
\author{Les Gasser}%
\affiliation{
Graduate School of Library and Information Science, and\\
Department of Computer Science,\\
University of Illinois at Urbana-Champaign.
}

\date{\today}

\begin{abstract}
Many important real-world networks manifest ``small-world" properties such as scale-free degree distributions, small diameters, and clustering.  The most common model of growth for these networks is ``preferential attachment", where nodes acquire new links with probability proportional to the number of links they already have. We show that preferential attachment is a special case of the process of molecular evolution. We present a new single-parameter model of network growth that unifies varieties of preferential attachment with the quasispecies equation (which models molecular evolution), and also with the Erd\~{o}s-R\'{e}nyi random graph model. We suggest some properties of evolutionary models that might be applied to the study of networks. We also derive the form of the degree distribution resulting from our algorithm, and we show through simulations that the process also models aspects of network growth. The unification allows mathematical machinery developed for evolutionary dynamics to be applied in the study of network dynamics, and vice versa.
\end{abstract}

%Use showkeys class option if keyword
\maketitle

\section{Introduction}
\label{introduction}

The study of networks has become a very active area of research since the discovery of ``small-world" networks \cite{milgram67smallWorld,
watts98smallWorld}.  Small-world networks are characterized by scale-free degree distributions, small diameters, and high clustering coefficients.
Many real networks, such as neuronal networks \cite{watts98smallWorld}, power grids \cite{asavathiratham01influenceModel}, the world wide web
\cite{albert99www} and human language \cite{ferrer-i-cancho01smallWorldOfLanguage}, have been shown to be small-world. Small-worldness has important
consequences. For example, such networks are found to be resistant to random attacks, but susceptible to targeted attacks, because of the power-law
nature of the degree distribution.

The process most commonly invoked for the generation of such networks is called ``preferential attachment'' \cite{barabasi99preferential,
albert00localEventsUniversality}. Briefly, new links attach preferentially to nodes with more existing links. Simon analyzed this stochastic process,
and derived the resulting distribution \cite{simon55skewDistribution}. This simple process has been shown to generate networks with many of the
characteristics of small-world networks, and has largely replaced the Erd\~{o}s-R\'{e}nyi random graph model \cite{erdos59randomGraphs} in modeling
and simulation work.

Another major area of research in recent years has been the consolidation of evolutionary dynamics \cite{page02unifying}, and its application to
alternate areas of research, such as language \cite{nowak02fromQuasispecies}. This work rests on the foundation of quasi-species theory
\cite{eigen77hypercycle, eigen88quasispecies}, which forms the basis of much subsequent mathematical modeling in theoretical biology.

In this paper we bring together network generation models and evolutionary dynamics models (and particularly quasi-species theory) by showing that
they have a common underlying probabilistic model. This unified model relates both processes through a single parameter, called a \emph{transfer
matrix}. The unification allows mathematical machinery developed for evolutionary dynamics to be applied in the study of network dynamics, and vice
versa. The rest of this paper is organized as follows: first we describe the preferential attachment algorithm and the quasispecies model of
evolutionary dynamics.  Then we show that we can describe both of these with a single probabilistic model. This is followed by a brief analysis, and
some simulations, which show that power-law degree distributions can be generated by   the model, and that the process can also be used to model some
aspects of network growth, such as densification power laws and shrinking diameters.

\section{Preferential Attachment}
\label{preferential attachment}
The Preferential Attachment algorithm specifies a process of network growth in which the addition of new (in-)links to nodes is random, but biased according to the number of (in-)links the node already has.  We identify each node by a unique type $i$, and let $x_i$ indicate the proportion of the total number of links in the graph that is already assigned to node $i$. Then equation \ref{preferential attachment equation} gives the probablity $P(i)$ of adding a new link to node $i$ \cite{barabasi99preferential}.
\begin{equation}
P(i) = \alpha x_{i}^{\gamma}.
\label{preferential attachment equation}
\end{equation}
where $\alpha$ is a normalizing term, and $\gamma$ is a constant. As $\gamma$ approaches $0$ the preference bias disappears; $\gamma>1$ causes exponentially greater bias from the existing in-degree of the node.

\section{Evolutionary Dynamics and Quasispecies}
\label{evolution and RME} Evolutionary dynamics describes a population of \emph{types} (species, for example) undergoing change through replication,
mutation, and selection\footnote{Simon (and Yule \cite{yule25evolution} before him) applied their stochastic model to the estimation of numbers of
species within genera, but the notion of quasi-species was unknown at the time, and it addresses a much wider range of issues than species
frequency.}. Suppose there are $N$ possible types, and let $s_{i,t}$ denote the number of individuals of type $i$ in the population at time $t$. Each
type has a fitness, $f_{i}$ which determines its probability of reproduction. At each time step, we select, with probability proportional to fitness,
one individual for reproduction. Reproduction is noisy, however, and there is a probability $q_{ij}$ that an  individual of type $j$ will generate an
individual of type $i$. The expected value of the change in the number of individuals of type $i$ at time $t$ is given by,
\begin{equation}
\label{quasispecies-equation}
\Delta s_{i,t} = \frac{\sum_{j}f_{j}s_{j}q_{ij}}{\sum_{j}f_{j}s_{j}}
\end{equation}
This is known as the quasispecies equation \cite{eigen88quasispecies}. The fitness, $f_{i}$, is a constant for each $i$. Fitness can also be \emph{frequency-dependent}, i.e. it can depend on which other types are present in the population. In this case the above equation is known as the replicator-mutator equation (RME) \cite{page02unifying},\cite{komarova04RME}.

\section{A Generalized Polya's Urn Model That Describes Both Processes}
\label{polya's urn}
Urn models have been used to describe both preferential attachment \cite{chung03polya}, and evolutionary processes \cite{benaim04evolutionaryUrnModels}. Here we describe an urn process derived from the quasispecies equation that also gives a model of network generation. In addition, this model of network generation will be seen to unify the Erd\~{o}s-R\'{e}nyi random graph model \cite{erdos59randomGraphs} with the preferential attachment model.

Our urn process is as follows:
\begin{itemize}
    \item We have a set of $n$ urns, which are all initially empty except for one, which has one ball in it.
    \item We add balls one by one, and a ball goes into urn $i$ with probability proportional to $f_{i}m_{i}$, where $f_{i}$ is the ``fitness" of urn $i$, and $m_{i}$ is the number of balls already in urn $i$.
    \item If the ball is put into urn $j$, then a ball is taken out of urn $j$, and moved to urn $k$ with probability $q_{kj}$.
\end{itemize}
The matrix $Q = [q_{ij}]$, which we call the \emph{transfer matrix}, is the same as the mutation matrix in the quasispecies equation.

This process describes the preferential attachment model if we set the fitness, $f_{i}$, to be proportional to $m_{i}^{\gamma-1}$, where $\gamma$ is a constant (as in equation \ref{preferential attachment equation}). Now we get a network generation algorithm in much the same way as Chung et al. did \cite{chung03polya}, where each ball corresponds to a half-edge, and each urn corresponds to a node. Placing a ball in an urn corresponds to linking to a node, and moving a  ball from one urn to another corresponds to rewiring. We call this algorithm \textit{Noisy Preferential Attachment} (NPA). If the transfer matrix is set to be the identity matrix, Noisy Preferential Attachment reduces to pure preferential attachment.

In the NPA algorithm, just like in the preferential attachment algorithm, the probability of linking to a node depends only on the number of in-links to that node. The ``from" node for a new edge is chosen uniformly randomly. In keeping with standard practice, the graphs in the next section show only the in-degree distribution. However, since the ``from" nodes are chosen uniformly randomly, the total degree distribution has the same form.
Consider the case where the transfer matrix is almost diagonal, i.e. $q_{ii}$ is close to 1, and the same $\forall i$, and all the $q_{ij}$ are small and equal, $\forall i \neq j$. Let $q_{ii} = p$ and
\begin{equation}
q_{ij} = \frac{1-p}{n-1} = q,   \forall i \neq j.
\end{equation}
Then, the probability of the new ball being placed in bin $i$ is
\begin{equation}
P(i) = \alpha m_{i}^{\gamma}p + (1-\alpha m_{i}^{\gamma})q,
\end{equation}
where $\alpha$ is a normalizing constant. That is, the ball could be placed in bin $i$ with probability $\alpha m_{i}^{\gamma}$ and then replaced in
bin $i$ with probability $p$, or it could be placed in some other bin with probability $(1-\alpha m_{i}^{\gamma})$, and then transferred to bin $i$
with probability $q$. Rearranging, we get,
\begin{equation}
P(i) = \alpha m_{i}^{\gamma}(p-q) + q.
\end{equation}
In this case, NPA reduces to preferential attachment with initial attractiveness \cite{dorogovtsev00preferentialLinking}, where the initial
attractiveness ($q$, here) is the same for each node. We can get different values of initial attractiveness by setting the transfer matrix to be
non-uniform. We can get the Erd\~{o}s-R\'{e}nyi model by setting the transfer matrix to be entirely uniform, i.e. $q_{ij} = 1/n, \forall i, j$. Thus
the Erd\~{o}s-R\'{e}nyi model and the preferential attachment model are seen as two extremes of the same process, which differ with the transfer
matrix, $Q$.

This process also obviously describes the evolutionary process when $\gamma=1$. In this case, we can assume that at each step we first select a ball
from among all the balls in all the urns with probability proportional to the fitness of the ball (assuming that the fitness of a ball is the same as
the fitness of the urn in which it is). The probability that we will choose a ball from urn $i$ is proportional to $f_{i}m_{i}$. We then replace this
ball and add  another ball to the same urn. This is the replication step. This is followed by a mutation step as before, where we choose a ball from
the urn and either replace it in the urn with with probability $p$ or move it to any one of the remaining urns. If we assume that all urns (i.e. all
types or species) have the same \emph{intrinsic} fitness, then this process reduces to the preferential attachment process.

Having developed the unified NPA model, we can now point towards several concepts in quasi-species theory that are missing from the study of
networks, that NPA makes it possible to investigate:

\begin{itemize}
    \item Quasi-species theory assumes a genome, a bit string for example. This allows the use of a distance measure on the space of types.
    \item Mutations are often assumed to be point mutations, i.e. they can flip one bit. This means that a mutation cannot result in just \emph{any} type being introduced into the population, only a neighbor of the type that gets mutated.
    \item This leads to the notion of a quasi-species, which is a cloud of mutants that are close to the most-fit type in genome space.
    \item Quasi-species theory also assumes a fitness landscape. This may in fact be flat, leading to neutral evolution \cite{kimura83neutral}. Another (toy) fitness landscape is the Sharply Peaked Landscape (SPL), which has only one peak and therefore does not suffer from problems of local optima. In general, though, fitness landscapes have many peaks, and the ruggedness of the landscape (and how to evaluate it) is an important concept in evolutionary theory. The notion of (node) fitness is largely missing from network theory (with a couple of exceptions: \cite{caldarelli02vertexIntrinsicFitness}, \cite{barabasi01multiscaling}), though the study of networks might benefit  greatly from it.
    \item The event of a new type entering the population and ``taking over" is known as fixation. This means that the entire population eventually consists of this new type. Typically we speak of gene fixation, i.e. the probability that a single new gene gets incorporated into all genomes present in the population. Fixation can occur due to drift (neutral evolution) as well as due to selection.
\end{itemize}

\section{Analysis and Simulations}
\label{analysis and simulations}

We next derive the degree distribution of the network. Since there is no ``link death" in the NPA algorithm and the number of nodes is finite, the
limiting behavior in our model is not the same as that of the preferential attachment model (which allows introduction of new nodes). This means that
we cannot re-use Simon's result \cite{simon55skewDistribution} directly to derive the degree distribution of the network that results from NPA.
\subsection{Derivation of the degree distribution}

Suppose there are $N$ urns and $n$ balls at time $t$. Let $x_{i,t}$ denote the fraction of urns with $i$ balls at time $t$. We choose a ball
uniformly at random and ``replicate" it, i.e. we add a new ball (and replace the chosen ball) into the same urn. Uniformly random choice corresponds
to a model where all the urns have equal intrinsic fitness. We follow this up by drawing another ball from this urn and moving it to a uniformly
randomly chosen urn (from the $N-1$ other urns) with probability $q=(1-p)/(N-1)$, where $p$ is the probability of putting it back in the same urn.
Let $P_{1}(i)$ be the probability that the ball to be replicated is chosen from an urn with $i$ balls. Let $P_{2}(i)$ be the probability that the new
ball is placed in an urn with $i$ balls. The net probability that the new ball ends up in an urn with $i$ balls,
\begin{equation}
P(i) = P_{1}(i)~\text{and}~P_{2}(i)~\text{or}~\bar{P}_{1}(i)~\text{and}~P_{2}(i).
\end{equation}
%where $P_{k}(i)$ is the probability that the new ball goes into an urn with $i$ balls in the $k^{th}$ step.
The probability of selecting a ball from an urn with $i$ balls,
\begin{equation*}
P_{1}(i) = \frac{Nx_{i,t}i}{n_{0}+t},
\end{equation*}
where $n_{0}$ is the number of balls in the urns initially. $P_{2}(i)$ depends on the outcome of the first step.
\begin{equation*}
P_{2}(i)=
\begin{cases}
p + (Nx_{i,t}-1)q & \text{when step 1 is ``successful"}, \\
Nx_{i,t}q & \text{when step 1 is a ``failure"}.
\end{cases}
\end{equation*}
Putting these together, we get,
\begin{eqnarray*}
P(i) &=& \frac{Nx_{i,t}i}{n_{0}+t}(p + (Nx_{i,t}-1)q) + \big(1-\frac{Nx_{i,t}i}{n_{0}+t}\big)Nx_{i,t}q \\
    &=& \frac{Nx_{i,t}i}{n_{0}+t}(p-q) + Nx_{i,t}q.
\end{eqnarray*}
Now we calculate the expected value of $x_{i,t+1}$. $x_{i,t}$ will increase if the ball goes into an urn with $i-1$ balls. Similarly it will decrease if the ball ends up in an urn with $i$ balls. Otherwise it will remain unchanged. Remembering that $x_{i,t}$ is the \emph{fraction} of urns with $i$ balls at time $t$, we write,
\begin{equation*}
Nx_{i,t+1} =
\begin{cases}
Nx_{i,t} + 1 & \text{w. p.  } \frac{Nx_{i-1,t}(i-1)}{n_{0}+t}(p-q) + Nx_{i-1,t}q, \\
Nx_{i,t} - 1 & \text{w. p.  } \frac{Nx_{i,t}i}{n_{0}+t}(p-q) + Nx_{i,t}q, \\
Nx_{i,t} & \text{otherwise.}
\end{cases}
\end{equation*}
From this, the expected value of $x_{i,t+1}$ works out to be,
\begin{equation}
\label{xit-general-diff-eq}
x_{i,t+1} = \big[1-\frac{i(p-q)}{n_{0}+t}-q\big]x_{i,t} + \big[\frac{(i-1)(p-q)}{n_{0}+t}+q\big]x_{i-1,t}.
\end{equation}

\begin{figure}
\centering
\includegraphics [width=0.5\linewidth]{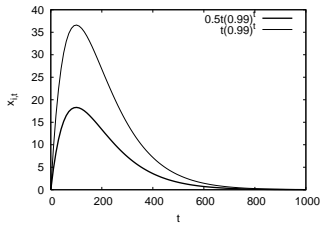}
\caption{Example $x_{i,t}$ curves.}
\label{xit-vs-t}
\end{figure}

We can show the approximate solution for $x_{i,t}$ to be,
\begin{equation}
\label{xit-solution-main} x_{i,t} = \frac{1-p}{N}\frac{r^{i-1}\Gamma(i)}{\prod_{k=1}^{i}(kr+1)}(t+1)(1-q)^{t-1},
\end{equation}
where $r=(p-q)/(1-q)$. This approximation is valid while $t << N$. See Appendix A for details. For any particular $i$, the shape of this curve is
given by $t(1-q)^t$. An example curve is shown in fig \ref{xit-vs-t}. This matches our intuition. Initially, $x_{i,t} = 0$ for $i>1$. As $t$
increases, $x_{i,t}$ increases through mutations. However, since $N$ is finite and we keep adding balls, eventually the number of bins with $i$ balls
must go to zero for any particular $i$. Thus $x_{i,t}$ must eventually start decreasing, which is what we see in figure \ref{xit-vs-t}. The middle
term can be simplified further as,
\begin{eqnarray*}
\frac{r^{i-1}}{\prod_{k=1}^{i}(kr+1)} &=& \frac{r^{i-1}}{\prod_{k=1+1/r}^{i+1/r}(kr)} \\
 &=& \frac{1}{r\prod_{k=1+1/r}^{i+1/r}(k)} \\
 &=& \frac{\Gamma(1/r)}{r^2\Gamma(i+1+1/r)}.
\end{eqnarray*}

\begin{figure}
\centering
\includegraphics [width=0.5\linewidth]{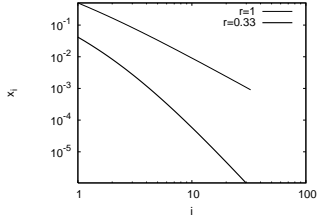}
\caption{The form of the degree distribution.}
\label{xit-vs-i}
\end{figure}

Therefore, in terms of $i$, equation \ref{xit-solution-main} can be written as (for fixed $t$),
\begin{equation}
\label{degree-distr-form}
x_{i} = C\frac{\Gamma(i)}{\Gamma(i+1+\frac{1}{r})},
\end{equation}
where $C$ is a constant. This is the form of the degree distribution. This is a power law, because as $i \rightarrow \infty$, equation
\ref{degree-distr-form} tends to $i^{-(1+1/r)}$ (see discussion of eq. 1.4 in \cite[ pg 426]{simon55skewDistribution}). This is also demonstrated in
the sample plots in figure \ref{xit-vs-i}.

\begin{figure}
\centering
\includegraphics [width=0.5\linewidth]{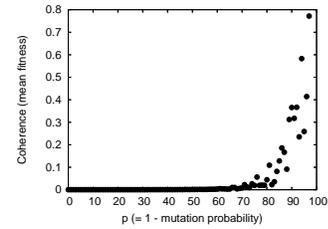}
\caption{N = 100000, number of edges = 10000.}
\label{coherence}
\end{figure}

\begin{figure}
\centering
\includegraphics [width=0.5\linewidth]{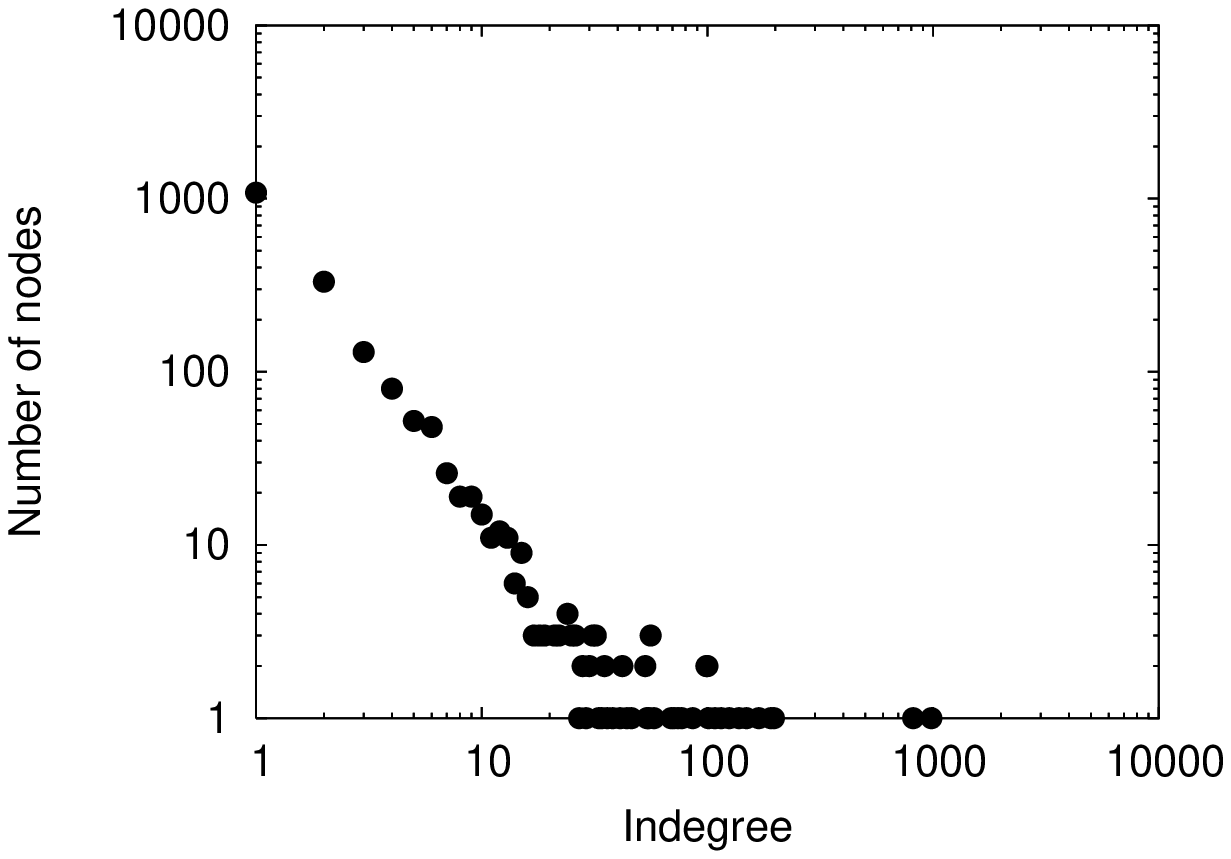}
\caption{p = 0.8, N = 100000, number of edges = 10000.}
\label{indegDist0.80-10000}
\end{figure}

\begin{figure}
\centering
\includegraphics [width=0.5\linewidth]{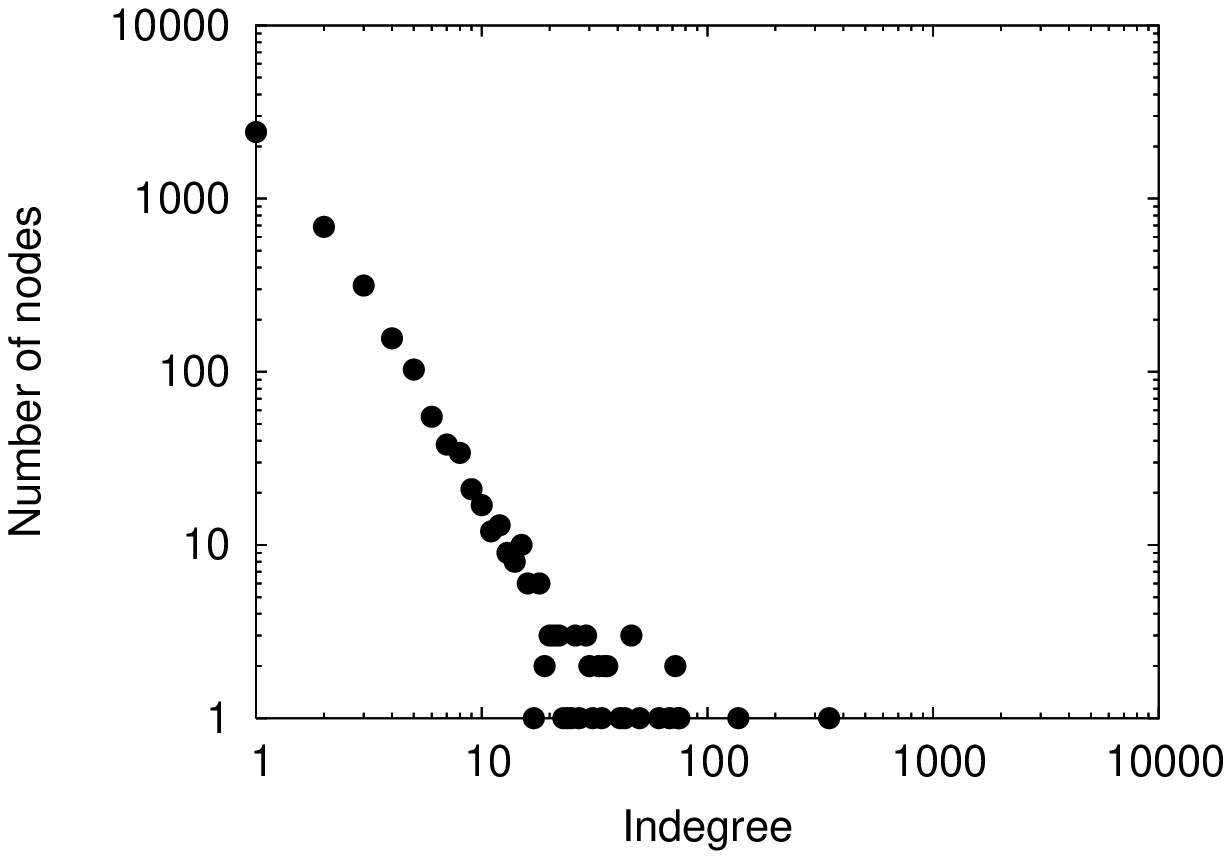}
\caption{p = 0.6, N = 100000, number of edges = 10000.}
\label{indegDist0.60-10000}
\end{figure}

\begin{figure}
\centering
\includegraphics [width=0.5\linewidth]{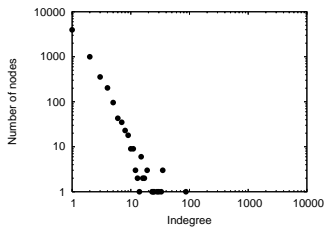}
\caption{p = 0.4, N = 100000, number of edges = 10000.}
\label{indegDist0.40-10000}
\end{figure}

These results are confirmed through simulation. We did an experiment where the number of possible nodes was set to $100000$, and $10000$ links were
added. The experiment was repeated for values of $p$ ranging from $0.01$ to $0.99$, in steps of $0.01$. Figure \ref{coherence} shows a plot of
\emph{coherence}, $\phi$, which is defined as,
\begin{equation}
\phi = \sum_{i}x_{i}^{2}.
\end{equation}
Coherence is a measure of the non-uniformity of the degree distribution. It is $1$ when a single node has all the links. When all nodes have one link
each, coherence has its lowest value, $1/N$. We see that as $p$ increases (i.e. mutation rate decreases), coherence also increases. This is borne out
by the degree distribution plots (figures \ref{indegDist0.80-10000} through \ref{indegDist0.40-10000}). The degree distribution is steeper for lower
values of $p$.

\subsection{Stability}
We can rewrite equation \ref{quasispecies-equation} as
\begin{equation}
\Delta s_{i} = \frac{1}{\sum_j f_{j} s_{j}}(f_{i}s_{i}q_{ii} + \sum_{j\neq i} f_{j}s_{j}q_{ij})
\end{equation}
The first term in the parentheses represents the change in $s_{i}$ due to selection. Some of the copies of type $i$ are lost due to mutation. The
fraction that are retained are given by the product $f_{i}q_{ii}$. If this product is greater than 1, the proportion of type $i$ will increase due to
selection, otherwise it will decrease. The second term represents the contribution to type $i$ due to mutation from all the other types in the
population. Thus, if $s_{i}$  decreases towards zero due to a selective disadvantage, it will be maintained in the population at ``noise" level due
to mutations.

%$N$, the size of the space of types, is generally extremely large. For example, if we have genomes of length 100 constructed from four base types,
%the space of all possible genomes (i.e. $N$) is $4^{100}$. This is a number large enough that the actual population will never sample all the
%possible genomes. In this case a reasonable assumption is that every mutation introduces a new type into the population. This means that the second
%term can be neglected and the persistence of type $i$ in the population essentially depends entirely on the mutation probability for type $i$. If
%$q_{ii}$ is too low, $s_{i}$ will decrease and eventually disappear from the population.

This leads to the notion of an error threshold. Suppose that the fitness landscape has only one peak. This is known as the Sharply Peaked Landscape,
or SPL. Suppose further that mutations only alter one position on the genome at a time. Then it can be shown that if the mutation rate is small
enough the population will be closely clustered about the fittest type. The fittest type keeps getting regenerated due to selection, and mutations
generate a cloud of individuals with genomes very close to the genome of the fittest type. This cloud is known as a \emph{quasi-species}
\cite{eigen89quasispecies}.

If, on the other hand, the mutation rate is above a certain threshold (essentially $1/f_{i}$, where $i$ is the fittest type) then all types will
persist in the population in equal proportions. This threshold is known as the error threshold.

\section{Fitness Landscapes and Neutral Evolution}

We have seen above that noisy preferential attachment is equivalent to molecular evolution where all intrinsic fitnesses are equal. If node fitnesses are allowed to be different, we get standard quasi-species behavior. If the mutation rate is low enough, the fittest node dominates the network and acquires nearly all the links. If the mutation rate is high enough to be over the error threshold, no single node dominates.

\begin{figure}
\centering
\includegraphics [width=0.5\linewidth]{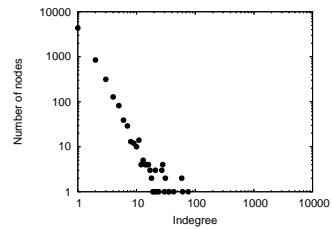}
\caption{p = 0.4, N = 100000, number of edges = 10000, node fitnesses are uniformly randomly distributed between 0 and 1.}
\label{indegDist0.4-10000-withFitness}
\end{figure}

\begin{figure}
\centering
\includegraphics [width=0.5\linewidth]{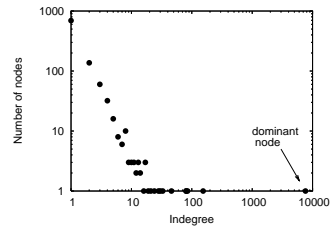}
\caption{p = 0.9, N = 100000, number of edges = 10000, node fitnesses are uniformly randomly distributed between 0 and 1.}
\label{indegDist0.9-10000-withFitness}
\end{figure}

Figures \ref{indegDist0.4-10000-withFitness} and \ref{indegDist0.9-10000-withFitness} show simulations where nodes are assigned intrinsic fitness values uniformly randomly in the range $(0,1)$, for different values of $p$. We see that when $p$ is high (0.9), i.e. mutation rate is low, the degree distribution stretches out along the bottom, and one or a few nodes acquire nearly all the links. When $p=0.4$, though, we don't get this behavior, because the mutation rate is over the error threshold.

Since we generally don't see a single node dominating in real-world networks, we are led to one of two
conclusions: either mutation rates in real-world networks are rather high, or the intrinsic fitnesses of the nodes are all equal. The former seems somewhat untenable. The latter suggests that most networks undergo neutral evolution \cite{kimura83neutral}.

\begin{figure}
\centering
\includegraphics [width=0.7\linewidth]{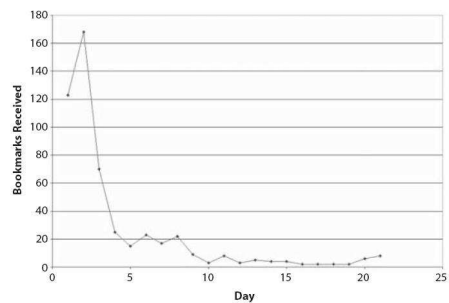}
\caption{This is figure 6a from \cite{huberman06tagging}. It shows number of bookmarks received against time (day number). This particular site
acquires a lot of bookmarks almost immediately after it appears, but thereafter receives few bookmarks.} \label{huberman6a}
\end{figure}

\begin{figure}
\centering
\includegraphics [width=0.7\linewidth]{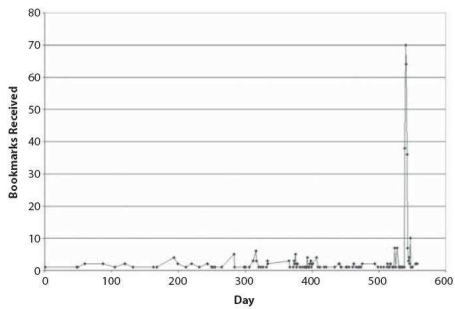}
\caption{This is figure 6b from \cite{huberman06tagging}. It shows number of bookmarks received against time (day number). This particular site
suddenly acquires a lot of bookmarks in a short period of time, though it has existed for a long time.} \label{huberman6b}
\end{figure}

Fitness landscapes can also be dynamic. Golder and Huberman give examples of short term dynamics in collaborative tagging systems (in particular
Del.icio.us) \cite{huberman06tagging}. Figures \ref{huberman6a} and \ref{huberman6b}, which are taken from their paper, show two instances of the
rate at which two different web sites acquired bookmarks. The first one shows a peak right after it appears, before the rate of bookmarking drops to
a baseline level. The second instance shows a web site existing for a while before it suddenly shows a peak in the rate of bookmarking. Both are
examples of dynamic, i.e. changing, fitness. Wilke et al. have shown that in the case of molecular evolution a rapidly changing fitness landscape is
equivalent to the time-averaged fitness landscape \cite{wilke01dynamicFitness}. Thus while short term dynamics show peaks in link (or bookmark)
acquisition, the long-term dynamics could still be neutral or nearly neutral.

\section{Dynamical properties of real-world networks}

Leskovec et al. point out that though models like preferential attachment are good at generating networks that match static ``snapshots" of
real-world networks, they do not appropriately model how real-world networks change over time \cite{kleinberg05graphsOverTime}. They point out two
main properties which are observed for several real-world networks over time: \emph{densification power laws}, and \emph{shrinking diameters}. The
term densification power law refers to the fact that the number of edges grows super-linearly with respect to the number of nodes in the network. In
particular, it grows as a power law. This means that these networks are getting more densely connected over time. The second surprising property of
the dynamics of growing real-world networks is that the diameter (or 90th percentile distance, which is called the \emph{effective} diameter)
\emph{decreases} over time. In most existing models of scale-free network generation, it has been shown that the diameter increases very slowly over
time \cite{bollobas04diameter}. Leskovec et al. stress the importance of modeling these dynamical aspects of network growth, and they present an
alternate algorithm that displays both the above properties.

\begin{figure}
\centering
\includegraphics [width=0.5\linewidth]{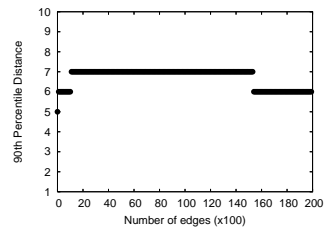}
\caption{The effective diameter of the network when the mutation rate decreases over time from 0.3 to 0.01. It increases quickly at first and then decreases slowly over time.}
\label{90thPercentileDistances}
\end{figure}

\begin{figure}
\centering
\includegraphics [width=0.5\linewidth]{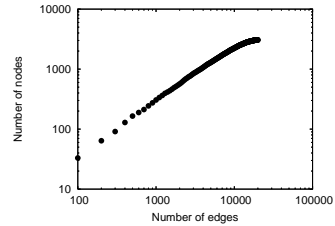}
\caption{The number of nodes grows as a power law with respect to the number of edges (or time, since one edge is added at each time step). The slope of the line is approximately 0.86.}
\label{nonzeroDegNodes}
\end{figure}

Noisy preferential attachment can also show these properties if we slowly decrease the mutation rate over time. Figures \ref{90thPercentileDistances} and \ref{nonzeroDegNodes} show the effective diameter of the network and the rate of change of the number of nodes with respect to the number of edges for a simulation in which the mutation rate was changed from 0.3 to 0.01 over the course of the simulation run.

\section{Conclusions}
\label{conclusions}

We have shown that, when modeled appropriately, the preferential attachment model of network generation can be seen as a special case of the process of molecular evolution because they share a common underlying probabilistic model. We have presented a new, more general, model of network generation, based on this underlying probabilistic model. Further, this new model of network generation, which we call Noisy Preferential Attachment, unifies the Erd\~{o}s-R\'{e}nyi random graph model with the preferential attachment model.

The preferential attachment algorithm assumes that the fitness of a node depends only on the number of links it has. This is not true of most real networks. On the world wide web, for instance, the likelihood of linking to an existing webpage depends also on the content of that webpage. Some websites also experience sudden spurts of popularity, after which they may cease to acquire new links. Thus the probability of acquiring new links depends on more than the existing degree. This kind of behavior can be modeled by the Noisy Preferential Attachment algorithm by including intrinsic fitness values for nodes.

The Noisy Preferential Attachment algorithm can also be used to model some dynamical aspects of network growth such as densification power laws and shrinking diameters by gradually decreasing mutation rate over time. If true, this brings up the intriguing question of why mutation rate would decrease over time in real-world networks. On the world wide web, for example, this may have to do with better quality information being available through the emergence of improved search engines etc. However, the fact that many different kinds of networks exhibit densification and shrinking diameters suggests that there may be some deeper explanation to be found.

From a design point of view, intentional modulation of the mutation rate can provide a useful means of trading off between exploration and exploitation of network structure. We have been exploring this in the context of convergence in a population of artificial language learners \cite{swarup06npaAndLangev}.

The larger contribution of this work, however, is to bring together the fields of study of networks and evolutionary dynamics, and we believe that many further connections can be made.

\section{Acknowledgements}
\label{acknowledgements}

We appreciate the helpful comments of Roberto Aldunate and Jun Wang. Work supported under NSF Grant IIS-0340996.

\appendix
\section{}

Here we solve the difference equation,
\begin{equation}
\label{xit-general-diff-eq-appendix}
x_{i,t+1} = \big[1-\frac{i(p-q)}{n_{0}+t}-q\big]x_{i,t} + \big[\frac{(i-1)(p-q)}{n_{0}+t}+q\big]x_{i-1,t}.
\end{equation}
$x_{0,t}$ is a special case.
\begin{equation*}
Nx_{0,t+1} =
\begin{cases}
Nx_{0,t} - 1 & \text{w. p.  } Nx_{0,t}q, \\
Nx_{0,t} & \text{otherwise.}
\end{cases}
\end{equation*}
Expanding and simplifying as above, we get,
\begin{equation*}
x_{0,t+1} = (1-q)x_{0,t}.
\end{equation*}
The solution to this difference equation is simply,
\begin{equation}
\label{x0t-solution}
x_{0,t} = (1-q)^{t}x_{0,0},
\end{equation}
where $x_{0,0}=(N-1)/N$ is the initial value of the number of empty urns. Note that here, and henceforth, we are assuming that initially all the urns are empty except for one, which has one ball in it. Therefore $x_{1,0}=1$, and $x_{i,0}=0~~\forall i>1$. This also means that $n_{0}=1$. These conditions together specify the entire initial state of the system.

Equation \ref{xit-general-diff-eq-appendix} is difficult to solve directly, so we shall take the approach of finding the solution to $x_{1,t}$ and $x_{2,t}$ and then simply guessing the solution to $x_{i,t}$.

Substituting $i=1$ in equation \ref{xit-general-diff-eq} gives us,
\begin{equation*}
x_{1,t+1} = \big[1-\frac{(p-q)}{n_{0}+t}-q\big]x_{1,t} + qx_{0,t}.
\end{equation*}
Substituting the solution for $x_{0,t}$ from equation \ref{x0t-solution} gives us,
\begin{equation}
\label{x1t-diff-eq}
x_{1,t+1}=\big[1-\frac{(p-q)}{n_{0}+t}-q\big]x_{1,t} + q(1-q)^{t}x_{0,0}.
\end{equation}

The complete solution for $x_{1,t}$ is (see Appendix B),
\begin{equation}
\label{x1t-solution}
x_{1,t} = (1-q)^{t}\big[A(t+1) + \frac{B}{t^{\underline{r}}}\big],
\end{equation}
where $A=\frac{qx_{0,0}}{1+p-2q}$ and $B=\frac{2(p-q)}{(1+p-2q)N\Gamma(1-r)}$ are constants.
Let us now use this result to derive the solution for $x_{2,t}$. Substituting $i=2$ in equation \ref{xit-general-diff-eq-appendix}, we get,
\begin{equation*}
x_{2,t+1} = \big[1-\frac{2(p-q)}{n_{0}+t}-q\big]x_{2,t} + \big[\frac{p-q}{n_{0}+t} + q\big]x_{1,t}.
\end{equation*}
Substituting the solution for $x_{1,t}$ from equation \ref{x1t-solution} and replacing $n_{0}$ by 1 for convenience gives us,
\begin{align}
\label{x2t-diff-eq} x_{2,t+1} &= \big[1-\frac{2(p-q)}{1+t}-q\big]x_{2,t} + \notag\\
& (1-q)^{t}\big[A(t+1) + \frac{B}{t^{\underline{r}}}\big]\big[\frac{p-q}{1+t} + q\big].
\end{align}
The solution to this (after some work) turns out to be (see Appendix B),
\begin{align}
x_{2,t} &= (1-q)^{t}\big[A(t+1)\frac{r}{2r+1} + \frac{B}{t^{\underline{r}}} + \frac{D}{t^{\underline{2r}}}\big] \notag \\
& + \frac{q(1-q)^{t}}{1+p-2q}\big[A(t+1)\frac{2rt+t+2r}{2(2r+1)} + \frac{B}{t^{\underline{r}}}(t+2)\big]
\end{align}
In the above expression, compared to the first term, the remaining terms are negligible. To see this, consider that $B/t^{\underline{r}}$ can be at most $B$ (as $r \rightarrow 0$), and at least $B/t$ (as $r \rightarrow 1$). $B$ itself is less than $1/N$. Therefore the contribution of the second term is upper-bounded by $1/N$. A similar observation will hold for $D/t^{\underline{2r}}$. This is far less than the contribution due to the first term, since $A$ (which   is also close to $1/N$) is multiplied by $(t+1)$. The remaining terms are approximately of the form $t^{2}/N^{2}$ (and  higher $i$ will contain higher powers). We can ignore these as long as $t << N$. Thus, we can write the solution for $x_{2,t}$ approximately as,
\begin{eqnarray*}
x_{2,t} &=& \frac{Ar}{2r+1}(t+1)(1-q)^{t} \\
 &=& \frac{q}{1+p-2q}\frac{N-1}{N}(t+1)(1-q)^t \\
 &=& \frac{1-p}{N}\frac{r}{(r+1)(2r+1)}(t+1)(1-q)^{t-1}.
\end{eqnarray*}
We can continue on with $x_{3,t}$:
\begin{equation*}
x_{3,t+1} = \big[1-\frac{3(p-q)}{1+t}-q\big]x_{3,t} + \big[\frac{2(p-q)}{1+t} + q\big]x_{2,t}.
\end{equation*}
If we follow through with this as for $x_{2,t}$, we will see the 2 from the constant in the second term $(\frac{2p}{t+1})$ appear as a factor in the first term of the solution for $x_{3,t}$. In the general expression for the solution, this appears as $\Gamma(i)$. Therefore, we can guess the approximate expression for $x_{i,t}$ to be,
\begin{equation}
\label{xit-solution-appendix}
x_{i,t} = \frac{1-p}{N}\frac{r^{i-1}\Gamma(i)}{\prod_{k=1}^{i}(kr+1)}(t+1)(1-q)^{t-1},
\end{equation}
which is the same as equation \ref{xit-solution-main}

\section{}

Equation \ref{x1t-diff-eq} is,
\begin{equation*}
x_{1,t+1}=\big[1-\frac{(p-q)}{n_{0}+t}-q\big]x_{1,t} + q(1-q)^{t}x_{0,0}.
\end{equation*}
This equation is of the form $y(t+1)=p(t)y(t) + r(t)$. The general form of the solution is,
\begin{equation}
\label{x1t-general-form}
y(t)=u(t)\big[\sum\frac{r(t)}{Eu(t)} + C\big],
\end{equation}
where $u(t)$ is the solution of the homogeneous part of the above equation, i.e. $u(t+1)=p(t)u(t)$, and $E$ is the time-shift operator, i.e. $Eu(t) = u(t+1)$. Now, the homogeneous part of equation \ref{x1t-diff-eq} is,
\begin{eqnarray*}
u(t+1) &=& \big(1-q-\frac{p-q}{n_{0}+t}\big)u(t) \\
 &=& \big(\frac{(1-q)t + (1-q)n_{0} - (p-q)}{n_{0}+t}\big)u(t) \\
 &=& (1-q)\big(\frac{t+n_{0}-\frac{p-q}{1-q}}{t+n_{0}}\big)u(t).
\end{eqnarray*}
The solution to this difference equation is,
\begin{equation}
\label{x1t-homogeneous-soln}
u(t)=C(1-q)^{t}\frac{\Gamma(t+n_{0}-r)}{\Gamma(t+n_{0})},
\end{equation}
where $r=(p-q)/(1-q)$, $C$ is a constant, and $\Gamma(\cdot)$ is the gamma-function, which is a ``generalization" of the factorial to the complex
plane. It is defined recursively as $\Gamma(n+1)=n\Gamma(n)$. The derivation of equation \ref{x1t-homogeneous-soln} is given in Appendix C. From
equations \ref{x1t-diff-eq}, \ref{x1t-general-form}, and \ref{x1t-homogeneous-soln}, we get,
\begin{eqnarray*}
\lefteqn{x_{1,t} =} \\
 & & \!\!\!\!\!\!\!\!\!\!\! C(1-q)^{t}\frac{\Gamma(t+n_{0}-r)}{\Gamma(t+n_{0})}\big[\sum\frac{qx_{0,0}(1-q)^{t}\Gamma(t+1+n_{0})}{C(1-q)^{t+1}\Gamma(t+1+n_{0}-r)} \\
 & & + D_{1}\big] \\
 &=& \frac{C(1-q)^{t}}{(t+n_{0}-1)^{\underline{r}}}\big[\frac{qx_{0,0}}{C(1-q)}\sum(t+n_{0})^{\underline{r}} + D_{1}\big] \\
&&~~~~~~~~~~~~~\text{($t^{\underline{r}}$ is read as ``$t$ to the $r$ falling")} \\
 &=& \frac{q(1-q)^{t-1}x_{0,0}}{(t+n_{0}-1)^{\underline{r}}}\frac{(t+n_{0})^{\underline{r+1}}}{r+1} + \frac{D(1-q)^{t}}{(t+n_{0}-1)^{\underline{r}}} \\
&&~~~~~~~~~~~~~\text{(where $D=CD_{1}$ is another constant)} \\
 &=& \frac{q(1-q)^{t}x_{0,0}}{1+p-2q}\frac{\Gamma(t+n_{0}-r)}{\Gamma(t+n_{0})}\frac{\Gamma(t+n_{0}+1)}{\Gamma(t+n_{0}-r)} \\
 & & + \frac{D(1-q)^{t}}{(t+n_{0}-1)^{\underline{r}}} \\
 &=& \frac{q(1-q)^{t}x_{0,0}(t+n_{0})}{1+p-2q} + \frac{D(1-q)^{t}}{(t+n_{0}-1)^{\underline{r}}}.
\end{eqnarray*}
Let us evaluate the constant by applying the initial conditions $t=0$, $x_{0,0}=(N-1)/N$, $x_{1,0}=1/N$, and $n_{0}=1$. We get,
\begin{eqnarray*}
\frac{1}{N} &=& \frac{q\frac{N-1}{N}}{1+p-2q} + D\Gamma(1-r) \\
1 &=& \frac{q(N-1)}{1+p-2q} + ND\Gamma(1-r).
\end{eqnarray*}
\begin{equation}
\label{D}
\text{Therefore,}~~D = \frac{2(p-q)}{(1+p-2q)N\Gamma(1-r)}.
\end{equation}
This gives us the complete solution for $x_{1,t}$ as,
\begin{equation*}
x_{1,t} = (1-q)^{t}\big[A(t+1) + \frac{B}{t^{\underline{r}}}\big],
\end{equation*}
where $A=\frac{qx_{0,0}}{1+p-2q}$ and $B=D=\frac{2(p-q)}{(1+p-2q)N\Gamma(1-r)}$ are constants. This is the same as equation \ref{x1t-solution}.

\subsection{Solution to equation \ref{x2t-diff-eq}}
Equation \ref{x2t-diff-eq} is,
\begin{eqnarray*}
x_{2,t+1} &=& \big[1-\frac{2(p-q)}{1+t}-q\big]x_{2,t} \\
& & + (1-q)^{t}\big[A(t+1) + \frac{B}{t^{\underline{r}}}\big]\big[\frac{p-q}{1+t} + q\big].
\end{eqnarray*}
Again, this equation is of the form of equation \ref{x1t-general-form}. The solution to the homogeneous part in this case is,
\begin{equation}
\label{x2t-homogeneous-soln}
u(t) = C(1-q)^{t}\frac{\Gamma(t+1-\frac{2(p-q)}{1-q})}{\Gamma(t+1)}.
\end{equation}
This is found in exactly the same way as equation \ref{x1t-homogeneous-soln} (see Appendix B). Now, from equations \ref{x1t-general-form}, \ref{x2t-diff-eq}, and \ref{x2t-homogeneous-soln}, we get,
\begin{eqnarray*}
\lefteqn{x_{2,t} =} \\
 & & \!\!\!\!\! \frac{C(1-q)^{t}}{t^{\underline{2r}}}\big[\sum\frac{(1-q)^{t}(A(t+1)+\frac{B}{t^{\underline{r}}})(\frac{p-q}{t+1} + q)}{C(1-q)^{t+1}\frac{1}{(t+1)^{\underline{2r}}}} + D_{1}\big] \\
 &=& \frac{C(1-q)^{t}}{t^{\underline{2r}}}\big[\frac{1}{C(1-q)}\big[A(p-q)\sum(t+1)^{\underline{2r}} \\
 & & + Aq\sum(t+1)(t+1)^{\underline{2r}} + B(p-q)\sum\frac{(t+1)^{\underline{2r}}}{t^{\underline{r}}(t+1)} \\
 & & + Bq\sum\frac{(t+1)^{\underline{2r}}}{t^{\underline{r}}}\big] + D_{1}\big].
\end{eqnarray*}
Solving the summations (see Appendix C), we get,
\begin{eqnarray*}
\label{x2t-solution} x_{2,t} &=& \frac{C(1-q)^{t}}{t^{\underline{2r}}}\big[\frac{1}{C(1-q)}\big[\frac{A(p-q)(t+1)^{\underline{2r+1}}}{2r+1} \\
 & & + Aq\big(\frac{t(t+1)^{\underline{2r+1}}}{2r+1}- \frac{(t+1)^{\underline{2r+2}}}{(2r+1)(2r+2)}\big) \\
 & & + B(p-q)\frac{t^{\underline{2r}}}{rt^{\underline{r}}} + Bq\frac{(t+2)t^{\underline{2r}}}{(1+r)t^{\underline{r}}}\big] + D \big].
\end{eqnarray*}
Simplifying,
\begin{eqnarray*}
x_{2,t} &=& (1-q)^{t}\big[\frac{Ar(t+1)}{2r+1} + \frac{Aq(t+1)(2rt+t+2r)}{(1-q)(2r+1)(2r+2)} \\
 & & + \frac{B}{t^{\underline{r}}} + \frac{Bq(t+2)}{(1-q)(1+r)t^{\underline{r}}}\big] + \frac{D(1-q)^{t}}{t^{\underline{2r}}} . \\
 &=& (1-q)^{t}\big[A(t+1)\frac{r}{2r+1} + \frac{B}{t^{\underline{r}}} + \frac{D}{t^{\underline{2r}}}\big] \\
 & & + \frac{q(1-q)^{t}}{1+p-2q}\big[A(t+1)\frac{2rt+t+2r}{2(2r+1)} + \frac{B}{t^{\underline{r}}}(t+2)\big]
\end{eqnarray*}
This is the same as equation \ref{x2t-solution}.

\section{}

\subsection{Derivation of equation \ref{x1t-homogeneous-soln}}
Equation \ref{x1t-homogeneous-soln} is the solution to the following difference equation:
\begin{equation*}
u(t+1) = (1-q)\big(\frac{t+n_{0}-\frac{p-q}{1-q}}{t+n_{0}}\big)u(t).
\end{equation*}
Note that all the factors in this equation are positive. Taking log, we get,
\begin{eqnarray*}
\text{log}~u(t+1) &=& \text{log}\big((1-q)\big(\frac{t+n_{0}-r}{t+n_{0}}\big)\big) + \text{log}~u(t), \\
\Delta \text{log}~u(t) &=& \text{log}\big((1-q)\big(\frac{t+n_{0}-r}{t+n_{0}}\big)\big), \\
\text{log}~u(t) &=& \sum\big[\text{log}(1-q) + \text{log}(t+n_{0}-r) \\
 & & - \text{log}(t+n_{0})\big] + D.
\end{eqnarray*}
Remembering that $\sum a=ta$, and $\sum \text{log}(t+a)=\text{log}\Gamma(t+a)$, we get,
\begin{eqnarray*}
\text{log}~u(t) &=& t\text{log}(1-q) + \text{log}\Gamma(t+n_{0}-r) \\
 & & - \text{log}\Gamma(t+n_{0}) + D, \\
\text{Therefore,}~~ u(t) &=& C(1-q)^{t}\frac{\Gamma(t+n_{0}-r)}{\Gamma(t+n_{0})}.
\end{eqnarray*}
This is the same as equation \ref{x1t-homogeneous-soln}.
\subsection{Derivation of equation \ref{x2t-solution}}
Equation \ref{x2t-solution} is the solution to the following difference equation:
\begin{eqnarray*}
x_{2,t} &=& \frac{C(1-q)^{t}}{t^{\underline{2r}}}\big[\frac{1}{C(1-q)}\big[A(p-q)\sum(t+1)^{\underline{2r}} \\
 & & + Aq\sum(t+1)(t+1)^{\underline{2r}} + B(p-q)\sum\frac{(t+1)^{\underline{2r}}}{t^{\underline{r}}(t+1)} \\
 & & + Bq\sum\frac{(t+1)^{\underline{2r}}}{t^{\underline{r}}}\big] + D_{1}\big].
\end{eqnarray*}
We shall solve each of the summations individually. At several points, we will use the summation by parts formula,
\begin{equation}
\label{summation-by-parts}
\sum\big(Ey(t)\Delta z(t)\big)=y(t)z(t) - \sum\big(z(t)\Delta y(t)\big).
\end{equation}
The first summation term can be obtained directly:
\begin{equation}
\label{x2t-first-summation}
\sum(t+1)^{\underline{2r}} = \frac{(t+1)^{\underline{2r+1}}}{2r+1} + C_{1}.
\end{equation}
The second summation term can be obtained using the summation by parts formula. Let $Ey(t)=t+1$. Then $y(t)=t$, and $\Delta y(t)=1$. Let $\Delta z(t) = (t+1)^{\underline{2r}}$. Then $z(t) = \frac{(t+1)^{\underline{2r+1}}}{2r+1}$. We get,
\begin{equation*}
\sum(t+1)(t+1)^{\underline{2r}} = \frac{(t+1)(t+1)^{\underline{2r+1}}}{2r+1} - \sum\frac{(t+1)^{\underline{2r+1}}}{2r+1}.
\end{equation*}
\begin{equation}
\label{x2t-second-summation}
\sum(t+1)(t+1)^{\underline{2r}} = \frac{(t+1)(t+1)^{\underline{2r+1}}}{2r+1} - \frac{(t+1)^{\underline{2r+2}}}{(2r+1)(2r+2)} + C_{2}.
\end{equation}
Before proceeding, we pause to calculate $\sum(1/t^{\underline{r}})$. Note that,
\begin{eqnarray*}
\Delta\frac{1}{t^{\underline{r}}} &=& \frac{1}{(t+1)^{\underline{r}}} - \frac{1}{t^{\underline{r}}} \\
 &=& \frac{t+1-r}{(t+1)t^{\underline{r}}} - \frac{1}{t^{\underline{r}}} \\
 &=& \frac{-r}{(t+1)t^{\underline{r}}} \\
 \frac{t+1}{-r}\Delta\frac{1}{t^{\underline{r}}} &=& \frac{1}{t^{\underline{r}}}.
\end{eqnarray*}
Taking summation, we get,
\begin{equation*}
\sum\frac{1}{t^{\underline{r}}} = \frac{1}{-r}\sum\big(Et\Delta\frac{1}{t^{\underline{r}}}\big).
\end{equation*}
Using the summation by parts formula, we get,
\begin{eqnarray*}
\sum\frac{1}{t^{\underline{r}}} &=& \frac{1}{-r}\big(\frac{t}{t^{\underline{r}}} - \sum\frac{1}{t^{\underline{r}}}\big) \\
\big(1-\frac{1}{r}\big)\sum\frac{1}{t^{\underline{r}}} &=& \frac{-t}{rt^{\underline{r}}}
\end{eqnarray*}
\begin{equation}
\label{sum-1-over-t-to-r-falling}
\sum\frac{1}{t^{\underline{r}}} = \frac{t}{(1-r)t^{\underline{r}}}
\end{equation}
We now proceed to the third summation term in the difference equation for $x_{2,t}$.
\begin{equation*}
\sum\frac{(t+1)^{\underline{2r}}}{t^{\underline{r}}(t+1)} = \sum\frac{t^{\underline{2r-1}}}{t^{\underline{r}}}
\end{equation*}
We shall again use the summation by parts formula. Let $Ey(t) = t^{\underline{2r-1}}$. Therefore $y(t)=(t-1)^{\underline{2r-1}}$, and $\Delta y(t) = (2r-1)(t-1)^{\underline{2r-2}}$. Let $\Delta z(t) = 1/t^{\underline{r}}$. Therefore $z(t) = t/(1-r)t^{\underline{r}}$ (from equation \ref{sum-1-over-t-to-r-falling}). We get,
\begin{eqnarray*}
\sum\frac{t^{\underline{2r-1}}}{t^{\underline{r}}} &=& \frac{t(t-1)^{\underline{2r-1}}}{(1-r)t^{\underline{r}}} \\
 & & - \sum\frac{2r-1}{1-r}\frac{t(t-1)^{\underline{2r-2}}}{t^{\underline{r}}} \\
 &=& \frac{t(t-1)^{\underline{2r-1}}}{(1-r)t^{\underline{r}}} \\
 & & - \frac{2r-1}{1-r}\sum\frac{t^{\underline{2r-1}}}{t^{\underline{r}}} \\
 \big(1 + \frac{2r-1}{1-r}\big)\sum\frac{t^{\underline{2r-1}}}{t^{\underline{r}}} &=& \frac{t}{1-r}\frac{(t-1)^{\underline{2r-1}}}{t^{\underline{r}}} \\
\sum\frac{t^{\underline{2r-1}}}{t^{\underline{r}}} &=& \frac{t^{\underline{2r}}}{rt^{\underline{r}}}
\end{eqnarray*}
Therefore,
\begin{equation}
\label{x2t-third-summation}
\sum\frac{(t+1)^{\underline{2r}}}{t^{\underline{r}}(t+1)} = \frac{t^{\underline{2r}}}{rt^{\underline{r}}}
\end{equation}
The fourth summation term in the difference equation for $x_{2,t}$ is similar to the third one.
\begin{equation*}
\sum\frac{(t+1)^{\underline{2r}}}{t^{\underline{r}}} = \sum\frac{(t+1)^{\underline{2r}}}{t^{\underline{r}}(t+1)}(t+1)
\end{equation*}
Let $Ey(t)=(t+1)$. Then $y(t)=t$, and $\Delta y(t) = 1$. Let $\Delta z(t) = \sum\frac{(t+1)^{\underline{2r}}}{t^{\underline{r}}(t+1)}$. Then $z(t) = \frac{t^{\underline{2r}}}{rt^{\underline{r}}}$ (from equation \ref{x2t-third-summation}). Therefore, using the summation by parts rule, we get,
\begin{equation}
\label{step-one}
\sum\frac{(t+1)^{\underline{2r}}}{t^{\underline{r}}} = t\frac{t^{\underline{2r}}}{rt^{\underline{r}}} - \frac{1}{r}\sum\frac{t^{\underline{2r}}}{t^{\underline{r}}}
\end{equation}
Now,
\begin{eqnarray*}
\sum\frac{t^{\underline{2r}}}{t^{\underline{r}}} &=& \sum\frac{(t+1-2r)t^{\underline{2r-1}}}{t^{\underline{r}}} \\
 &=& \frac{(t-2r)t^{\underline{2r}}}{rt^{\underline{r}}} - \frac{1}{r}\sum\frac{t^{\underline{2r}}}{t^{\underline{r}}} \\
 &=& \frac{t-2r}{1+r}\frac{t^{\underline{2r}}}{t^{\underline{r}}}
\end{eqnarray*}
Substituting back in equation \ref{step-one}, we get,
\begin{eqnarray*}
\sum\frac{(t+1)^{\underline{2r}}}{t^{\underline{r}}} &=& t\frac{t^{\underline{2r}}}{rt^{\underline{r}}} - \frac{1}{r}\big(\frac{t-2r}{1+r}\frac{t^{\underline{2r}}}{t^{\underline{r}}}\big) \\
 &=& \frac{t^{\underline{2r}}}{rt^{\underline{r}}}\big(t-\frac{t-2r}{1+r}\big)
\end{eqnarray*}
Therefore, we have,
\begin{equation}
\label{x2t-fourth-summation}
\sum\frac{(t+1)^{\underline{2r}}}{t^{\underline{r}}} = \frac{(t+2)t^{\underline{2r}}}{(1+r)t^{\underline{r}}}
\end{equation}
Combining equations \ref{x2t-first-summation}, \ref{x2t-second-summation}, \ref{x2t-third-summation}, and \ref{x2t-fourth-summation}, we get the solution for $x_{2,t}$, i.e. equation \ref{x2t-solution}.

\bibliography{networks-and-evolution}

\end{document}